\documentclass[aps,prb,twocolumn,english,showpacs,superscriptaddress,amssymb,amsfonts]{revtex4}
\usepackage[T1]{fontenc}
\usepackage[latin9]{inputenc}
\usepackage{amsmath}
\usepackage{dsfont}
\usepackage{amstext}

\usepackage{tocvsec2}

\usepackage{amssymb}
\usepackage{amsbsy}
\usepackage{amsthm}
\usepackage{epsfig}
\usepackage{graphicx}
\usepackage{bbm}
\usepackage{hyperref}
\usepackage{color}
\usepackage{multirow}
\usepackage{pdfsync}
\def\beq{\begin{equation}}
\def\eeq{\end{equation}}

\newcommand{\cs}{{C_6}}

\newcommand{\Ref}[1]{Ref.~\onlinecite{#1}}

\newcommand{\bss}{{\boldsymbol{\sigma}}}
\newcommand{\bst}{{\boldsymbol{T}}}

\newcommand{\bse}{{\boldsymbol{e}}}
\newcommand{\bsg}{{\boldsymbol{g}}}

\newcommand{\ie}{{\emph{i.e.~}}}
\makeatletter

\newcommand{\Rmnum}[1]{\expandafter\@slowromancap\romannumeral #1@}
\makeatother
\newcommand{\imth}{\hspace{1pt}\mathrm{i}\hspace{1pt}}
\newcommand{\alert}[1]{{\color{red}{#1}}}
\newcommand{\blert}[1]{{\color{blue}{#1}}}
\newcommand{\eg}{{\emph{e.g.~}}}

\newcommand{\bea}{\begin{eqnarray}}
\newcommand{\eea}{\end{eqnarray}}
\newcommand{\bpm}{\begin{pmatrix}}
\newcommand{\epm}{\end{pmatrix}}
\newcommand{\bal}{\begin{aligned}}
\newcommand{\eal}{\end{aligned}}

\makeatother

\usepackage{babel}
\begin{document}
\title{Symmetric $Z_2$ spin liquids and their neighboring phases on triangular lattice}

\author{Yuan-Ming Lu}
%\author{Ashvin Vishwanath}
%\author{Dung-Hai Lee}
%\affiliation{Department of Physics, University of California, Berkeley, CA 94720, USA}
%\affiliation{Materials Science Division, Lawrence Berkeley National Laboratories, Berkeley, CA 94720}
\affiliation{Department of Physics, The Ohio State University, Columbus, OH 43210, USA}

\begin{abstract}
Motivated by recent numerical discovery of a gapped spin liquid phase in spin-$1/2$ triangular-lattice $J_1$-$J_2$ Heisenberg model, we classify symmetric $Z_2$ spin liquids on triangular lattice in the Abrikosov-fermion representation. We find 20 phases with distinct spinon symmetry quantum numbers, 8 of which have their counterparts in the Schwinger-boson representation. Among them we identify 2 promising candidates (\#1 and \#20), which can realize a gapped $Z_2$ spin liquid with up to next nearest neighbor mean-field amplitudes. We analyze their neighboring magnetic orders and valence bond solid patterns, and find one state (\#20) that is connected to 120-degree Neel order by a continuous quantum phase transition. We also identify gapped nematic $Z_2$ spin liquids in the neighborhood of the symmetric states and find 3 promising candidates (\#1, \#6 and \#20).
\end{abstract}


\maketitle

%{\small \setcounter{tocdepth}{2} \tableofcontents}

\section{Introduction}

Thanks to advances in numerical methods and increasing computational power, recently a lot of progress has been made in the pursuit of quantum spin liquids\cite{Balents2010,Lee2014a} (QSLs) in various frustrated quantum spin models\cite{Yan2011,Isakov2011,Jiang2012a,Depenbrock2012,Jiang2012,Wang2011a,Zhu2015,Hu2015,Rachel2015}. In contrast to classical paramagnets driven by thermal fluctuations, QSLs are symmetric zero-temperature phases featured by collective excitations which obey fractional statistics\cite{Kivelson1987,Read1989c,Kitaev2003} and long-range quantum entanglement\cite{Levin2006,Kitaev2006a} therein. Most recently using the density matrix renormalization group (DMRG) method, two numerical studies\cite{Zhu2015,Hu2015} of the spin$-1/2$ Heisenberg $J_1$-$J_2$ model on triangular lattice independently reported strong evidence for a gapped spin liquid phase\cite{Zhu2015,Hu2015} in the finite range of $J_2/J_1$, sandwiched by a noncollinear 120-degree Neel order at smaller $J_2/J_1$ and a collinear stripy Neel order at larger $J_2/J_1$. This is also supported by earlier variational Monte Carlo\cite{Sindzingre1995,Mishmash2013,Kaneko2014} and other\cite{Lecheminant1995,Jolicoeur1990,Li2015} studies on the same model. Due to its similarity to the gapped $Z_2$ spin liquid phase discovered in kagome Heisenberg model\cite{Yan2011,Jiang2012a,Depenbrock2012}, this discovery motivates us to look for symmetric $Z_2$ spin liquid phases on triangular lattice as possible candidates for the one realized in $J_1$-$J_2$ model.

A symmetric $Z_2$ spin liquid hosts 3 types of fractionalized excitations: bosonic spinon $b$, bosonic vison $v$ and fermionic spinon $f$. Each of them obeys mutual semionic statistics\cite{Kitaev2003} with the other two. Unlike usual ordered phases characterized by their broken symmetries, symmetric spin liquids are characterized by the symmetry quantum numbers of their fractional excitations. In particular, each quasiparticle can carry a fraction of the unit symmetry quantum number carried by a fundamental particle (such as a magnon in frustrated magnets), these phenomena are generally coined ``symmetry fractionalization''\cite{Essin2013,Mesaros2013}. Take $Z_2$ spin liquids for instance, each spinon carries spin-$1/2$, half of the spin quantum number carried by each magnon. Taken into account of time reversal, spin rotation and space group symmetries, there are many different symmetric $Z_2$ spin liquids with distinct qusiparticle symmetry quantum numbers. Instead of attempting to fully classify\cite{Essin2013,Mesaros2013,Song2015} all possible $Z_2$ spin liquids on triangular lattice, in this work we restrict ourselves to the Abrikosov-fermion representation\cite{Abrikosov1965,Affleck1988b,Wen2002}, which is a convenient way to construct variational wavefunctions of $Z_2$ spin liquids. Among 20 distinct symmetric states in Abrikosov-fermion representation, we identify two promising candidates (\#1 and \#20 in TABLE \ref{tab:PSG}) that can realize a gapped $Z_2$ spin liquid with up to next nearest neighbor mean-field amplitudes.

Since the gapped spin liquid appears in proximity to two magnetic orders in DMRG studies\cite{Zhu2015,Hu2015}, it's desirable to understand the symmetry-breaking phases in the neighborhood of symmetric $Z_2$ spin liquids. In particular, a $Z_2$ spin liquid can enter either a noncollinear magnetic order\cite{Sachdev1992,Chubukov1994a} or a valence bond solid (VBS) phase\cite{Read1989a,Moessner2001b} via continuous quantum phase transitions, by condensing either bosonic spinons $b$ or visons $v$. In this work, we study the magnetic and VBS phases in the neighborhood of the two gapped $Z_2$ spin liquids, \#1 and \#20 in TABLE \ref{tab:PSG}. In particular we find that \#20 can be driven into 120-degree noncollinear Neel order by a continuous quantum phase transition.

In DMRG studies of triangular $J_1$-$J_2$ model, strong evidence for nematic order \ie breaking of 6-fold rotational symmetry has been observed\cite{Zhu2015,Hu2015}. In this work we also explore the possibility of gapped nematic spin liquids\cite{Zhou2002} in the neighborhood of symmetric spin liquids. In addition to the 2 symmetric candidates (\#1 and \#20), we identify another promising nematic gapped state \ie \#6 in TABLE \ref{tab:PSG}. The mean-field ansatz of \#6 and \#20 nematic states are depicted in FIG. \ref{fig:6 nematic} and \ref{fig:20 nematic}.

This paper is organized as follows. In section \ref{CLASSIFICATION} we classify 20 distinct symmetric $Z_2$ spin liquids in the Abrikosov-fermion representation, as summarized in TABLE \ref{tab:PSG}. We also unify the 8 previously obtained $Z_2$ spin liquids in the Schwinger-boson representation\cite{Wang2006} with our Abrikosov-fermion states. In section \ref{SYMMETRIC Z2SL} we propose 2 promising candidates (\#1 and \#20 in TABLE \ref{tab:PSG}) among all 20 symmetric states for the $J_1$-$J_2$ spin liquid phase, and study the magnetic orders and VBS patterns in the neighborhood of these 2 promising candidates. In section \ref{NEMATIC Z2SL} we introduce nematic order into symmetric states, to achieve nematic $Z_2$ spin liquids on triangular lattice. We find 3 promising gapped nematic $Z_2$ spin liquids (\#1, \#6 and \#20 in TABLE \ref{tab:PSG}, see FIG. \ref{fig:20 nematic} and \ref{fig:6 nematic}) that may be realized in triangular-lattice $J_1$-$J_2$ model. Finally concluding remarks are given in section \ref{SUMMARY}.

\section{Classifying symmetric $Z_2$ spin liquids on triangular lattice}\label{CLASSIFICATION}

\begin{figure}
\includegraphics[width=0.8\columnwidth]{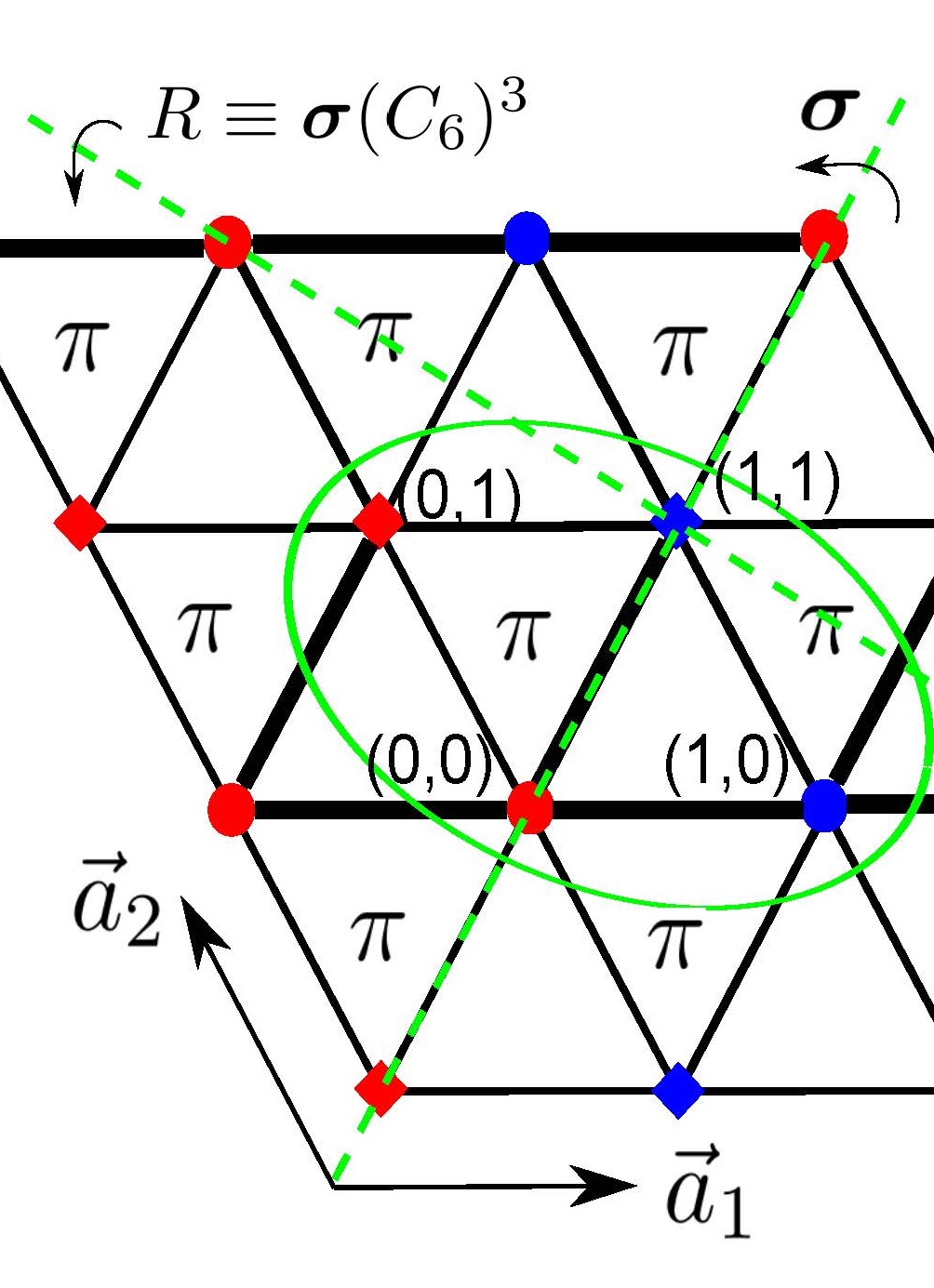}
\caption{(color online) Crystal symmetry of triangular lattice, and nearest-enighbor (NN) hopping amplitudes of fermionic spinons in the $\pi$-flux state on the triangle lattice. All NN hoppings are real with the same magnitude $|\alpha|>0$, where thick (thin) bonds have negative (positive) hopping signs. Each magnetic unit cell (denoted by the green oval) is chosen to contain 4 lattice sites labeled by red/blue diamonds/dots.}
\label{fig:triangle pi-flux}
\end{figure}

We label each lattice site on a triangular lattice by its positional vector ${\bf r}=x\vec{a}_1+y\vec{a}_2\equiv(x,y)$, where $\vec a_{1}=a(1,0)$ and $\vec a_2=a(-1,\sqrt3)/2$ are chosen to be two Bravais vectors as shown in FIG. \ref{fig:triangle pi-flux}. The crystal symmetry group is generated by two translations $T_{1,2}$, mirror reflection $\bss$ and site-centered $\pi/3$ rotation $\cs$
\bea
&(x,y)\overset{T_1}\longrightarrow(x+1,y),~~~(x,y)\overset{T_2}\longrightarrow(x,y+1),\notag\\
&(x,y)\overset{\bss}\longrightarrow(y,x),~~~(x,y)\overset{\cs}\longrightarrow(x-y,x).
\eea
These crystal symmetries, together with global time reversal $\bst$ and $SU(2)$ spin rotational symmetries impose constraints on a gapped $Z_2$ spin liquids. These symmetry conditions give rise to distinct $Z_2$ spin liquids in the Abrikosov-fermion representation.

\subsection{Classifying $Z_2$ spin liquids in Abrikosov-fermion representation}

In Abrikosov-fermion representation\cite{Abrikosov1965,Affleck1988b}, a spin $1/2$ on lattice site ${\bf r}$ is written in terms of two fermions $f_{{\bf r},\uparrow/\downarrow}$:
\bea
{\vec S}_{\bf r}=\frac14\text{Tr}\Big(\Psi^\dagger_{\bf r}\Psi_{\bf r}\vec\sigma\Big),~~~\Psi_{\bf r}\equiv\bpm f_{{\bf r},\uparrow}&f_{{\bf r},\downarrow}\\f_{{\bf r},\downarrow}^\dagger&-f^\dagger_{{\bf r},\uparrow}\epm
\eea
where $\vec\sigma$ stands for Pauli matrices. First we construct mean-field Hamiltonian (``ansatz'') for fermionic spinons
\bea
&H_{MF}=\sum_{{\bf r},{\bf l}}\text{Tr}\Big(\Psi^\dagger_{\bf r}\langle{\bf r}|{\bf l}\rangle\Psi_{\bf l}\Big),~~~~~\langle{\bf r}|{\bf l}\rangle=\langle{\bf l}|{\bf r}\rangle^\dagger.
\eea
where mean-field amplitudes $\langle{\bf r}|{\bf l}\rangle$ preserve all symmetries. Then the many-spin wavefunction can be obtained by Gutzwiller projection which enforces the following single-occupancy constraint
\bea
f^\dagger_{{\bf r},\uparrow}f_{{\bf r},\uparrow}+f_{{\bf r},\downarrow}^\dagger f_{{\bf r},\downarrow}=1
\eea
on the mean-field groundstate $|MF\rangle$ of fermionic spinons. There is a $SU(2)$ gauge redundancy\cite{Affleck1988b,Wen2002,Chen2012a} in this representation: spin operators $\vec S_{\bf r}$ remains invariant under \emph{local} $SU(2)$ gauge rotation $\Psi_{\bf r}\rightarrow W_{\bf r}\Psi_{\bf r},~W_{\bf r}\in SU(2)$. Therefore each physical symmetry operation $\hat u$ is followed by a $SU(2)$ gauge rotation $G_{\hat u}({\bf r})$ when acting on fermionic spinons. It follows that distinct $Z_2$ spin liquids are classified by so-called \emph{projective symmetry group} (PSG)\cite{Wen2002}: \ie gauge-inequivalent symmetry operations $\{G_{\hat u}({\bf r})|\hat u\in\text{symmetry~group}\}$ with $G_{\bse}({\bf r})=\pm1\in Z_2$ for the identity element $\bse$ of symmetry group.

In our case of triangular lattice, the symmetry group (including spin rotations, time reversal $\bst$ and space group generated by $T_{1,2},\bss,\cs$) imposes the algebraic conditions summarized on the left column of TABLE \ref{tab:unification} for symmetry operations $\{G_{\hat u}({\bf r})\in SU(2)\}$. The gauge-inequivalent solutions are the following:
\bea
&\notag G_{T_1}(x,y)=1,~~~G_{T_2}(x,y)=\eta_{12}^x,\\
&\notag G_\bss(x,y)=\eta_{12}^{xy}g_\bss,~~~G_{\cs}(x,y)=\eta_{12}^{xy+y(y-1)/2}g_{\cs},\\
&G_\bst(x,y)=g_\bst.
\eea
with
\bea\label{psg:simplify}
& g_\bss^2=\eta_\bss,~~g_\cs^6=\eta_\cs,~~(g_\cs g_\bss)^2=\eta_{\bss\cs},\\
&\notag g_\bst^2=\eta_\bst,~~g_\bss g_\bst=g_\bst g_\bss\eta_{\bss\bst},~~g_\cs g_\bst=g_\bst g_\cs\eta_{\cs\bst}.
\eea
Notice that we've defined the time reversal action $G_\bst({\bf r})$ on fermionic spinons $f_{{\bf r},\sigma}$ in the following way\cite{Wen2002}
\bea
\notag&\Phi_{\bf r}\overset{\bst}\longrightarrow G_\bst({\bf r})\tau^2\Phi_{\bf r}\sigma^y,~~~\Phi_{\bf r}\equiv\bpm f_{{\bf r},\uparrow}&f_{{\bf r},\downarrow}\\f_{{\bf r},\downarrow}^\dagger&-f^\dagger_{{\bf r},\uparrow}\epm.
\eea
In this notation, the mean-field amplitude $\langle{\bf r}_1|{\bf r}_2\rangle$ transforms as
\bea\label{mf condition:TRS}
\langle{\bf r}_1|{\bf r}_2\rangle=-G_{\bst}({\bf r}_1)\langle{\bf r}_1|{\bf r}_2\rangle G^\dagger_{\bst}({\bf r}_2).
\eea
under anti-unitary time reversal $\bst$ and
\bea\label{mf condition:unitary sym}
\langle{\bsg\bf r}_1|\bsg{\bf r}_2\rangle=G_{\bsg}(\bsg{\bf r}_1)\langle{\bf r}_1|{\bf r}_2\rangle G^\dagger_{\bsg}(\bsg{\bf r}_2).
\eea
under any (global and spatial) unitary symmetry $\bsg$.

If $\eta_\bst=1$ we have $g_\bst=1$, and hence all mean-field amplitudes must vanish identically due to time reversal symmetry (\ref{mf condition:TRS}). Therefore the only physical choice is $\eta_\bst=-1$ and we can choose a gauge so that $g_\bst=\imth\tau^2=G_\bst(x,y)$, which only allows real hopping ($\tau^3$) and real pairing ($\tau^1$) amplitudes. With this we summarize all symmetric PSGs in TABLE \ref{tab:PSG}. They correspond to 20 distinct symmetric $Z_2$ spin liquids on triangular lattice.

In addition to all symmetric states summarized in TABLE \ref{tab:PSG}, there is an extra pair of solutions to the PSG equations (\ref{psg:simplify}) with
\bea
&g_\bst=\imth\tau^2,~~g_\bss=\imth\tau^3,~~~g_\cs=e^{\imth\frac{n_6\pi}{6}\tau^2}~(n_6=1,2).\notag
\eea
as noticed in \Ref{Bieri2015}, In particular the $d+\imth d$ paired state of Abrikosov fermions proposed in \Ref{Mishmash2013} corresponds to the case of $n_6=2$. 

\begin{table*}[tb]
\centering
\begin{tabular} {|c||c|c|c||c|c|c|||c|c|c|c||c||}
\hline\hline
\multicolumn{7}{|c||}{Symmetric Abrikosov-fermion states}&\multicolumn{4}{|c|||}{Nematic states: mean-field amplitudes}&Schwinger-boson states\\
\hline
Label&$\eta_{12}$&$g_\bss$&$g_\cs$&
onsite $[0,0]$&NN $[1,1]$&NNN $[2,1]$&NN $[1,0]$&NN $[1,1]$&NNN $[2,1]$&NNN $[1,-1]$&$(p_1,p_2,p_3)$\\ \hline
\alert{\#1}&1&$\tau^0$&$\tau^0$&$\tau^{1,3}$&$\tau^{1,3}$&$\tau^{1,3}$&$\tau^{1,3}$&$\tau^{1,3}$&$\tau^{1,3}$&$\tau^{1,3}$&\alert{(1,1,0)}\\ \hline
\#2&-1&$\tau^0$&$\tau^0$&$\tau^{1,3}$&0&0&0&0&0&0&(0,1,0)\\ \hline
\#3&1&$\tau^0$&$\imth\tau^2$&0&0&0&0&0&0&0&\\ \hline
\blert{\#4}&-1&$\tau^0$&$\imth\tau^2$&0&0&$\tau^{1,3}$&$\tau^{1,3}$&0&$\tau^{1,3}$&$\tau^{1,3}$&\\ \hline
\#5&1&$\tau^0$&$\imth\tau^3$&$\tau^3$&$\tau^3$&$\tau^3$&$\tau^3$&$\tau^3$&$\tau^3$&$\tau^3$&(1,1,1)\\ \hline
\alert{\#6}&-1&$\tau^0$&$\imth\tau^3$&$\tau^3$&0&$\tau^1$&$\tau^1$&0&$\tau^1$&$\tau^1$&\alert{(0,1,1)}\\ \hline
\blert{\#7}&1&$\imth\tau^2$&$\tau^0$&0&0&0&$\tau^{1,3}$&0&$\tau^{1,3}$&0&\\ \hline
\#8&-1&$\imth\tau^2$&$\tau^0$&0&0&0&0&0&0&0&\\ \hline
\#9&1&$\imth\tau^2$&$\imth\tau^2$&0&0&0&0&0&0&0&\\ \hline
\blert{\#10}&-1&$\imth\tau^2$&$\imth\tau^2$&0&$\tau^{1,3}$&0&$\tau^{1,3}$&$\tau^{1,3}$&$\tau^{1,3}$&0&\\ \hline
\#11&1&$\imth\tau^2$&$\imth\tau^3$&0&0&0&$\tau^3$&0&$\tau^3$&0&\\ \hline
\#12&-1&$\imth\tau^2$&$\imth\tau^3$&0&$\tau^1$&0&$\tau^1$&$\tau^1$&$\tau^1$&0&\\ \hline
\blert{\#13}&1&$\imth\tau^3$&$\tau^0$&$\tau^3$&$\tau^3$&$\tau^3$&$\tau^{1,3}$&$\tau^3$&$\tau^{1,3}$&$\tau^3$&\blert{(1,0,0)}\\ \hline
\#14&-1&$\imth\tau^3$&$\tau^0$&$\tau^3$&0&0&0&0&0&0&(0,0,0)\\ \hline
\#15&1&$\imth\tau^3$&$\imth\tau^1$&0&0&0&$\tau^1$&0&$\tau^1$&0&\\ \hline
\#16&-1&$\imth\tau^3$&$\imth\tau^1$&0&0&$\tau^3$&$\tau^3$&0&$\tau^3$&$\tau^3$&\\ \hline
\#17&1&$\imth\tau^3$&$\imth\tau^2$&0&0&0&0&0&0&0&\\ \hline
\alert{\#18}&-1&$\imth\tau^3$&$\imth\tau^2$&0&$\tau^1$&$\tau^3$&$\tau^{1,3}$&$\tau^1$&$\tau^{1,3}$&$\tau^3$&\\ \hline
\#19&1&$\imth\tau^3$&$\imth\tau^3$&$\tau^3$&$\tau^3$&$\tau^3$&$\tau^3$&$\tau^3$&$\tau^3$&$\tau^3$&(1,0,1)\\ \hline
\alert{\#20}&-1&$\imth\tau^3$&$\imth\tau^3$&$\tau^3$&$\tau^1$&0&$\tau^1$&$\tau^1$&$\tau^1$&0&\alert{{(0,0,1)}}\\
\hline\hline
\end{tabular}
\caption{Summary of 20 distinct symmetric $Z_2$ spin liquids on triangular lattice. Among them, 8 states have their counterparts in the Schwinger-boson representation\cite{Wang2006}. We use $[x,y]\equiv\langle0,0|x,y\rangle$ to label representative mean-field amplitudes. Only 4 Abrikosov fermion states support a symmetric $Z_2$ spin liquid with up to next nearest neighbor (NNN) mean-field amplitudes, as highlighted by red color. Each symmetric Abrikosov-fermion state have a unique nematic descendant which breaks $C_6$ rotation but preserves mirror reflections $\bss$ and $R=\bss(\cs)^3$. In addition to the 4 symmetric ansatz, another 4 states support a nematic $Z_2$ spin liquid with up to NNN amplitudes, as highlighted by blue color. In the mean-field amplitudes, $\tau^1$ represents real pairing while $\tau^3$ stands for real hopping. Up to NNN mean-field amplitudes, only 3 ansatz support a gapped nematic $Z_2$ spin liquids: they are \#1, \#6 and \#20 states.}
\label{tab:PSG}
\end{table*}

%The symmetry-allowed mean-field amplitudes $[x,y]\equiv\langle0,0|x,y\rangle$ satisfy the following symmetry conditions
%\bea
%&\notag g_\bst[x,y]g^\dagger_\bst=-\eta_{1\bst}^{x+y}[x,y],\\
%&g_{\bp}g_P[x,y]g^\dagger_P g^\dagger_{\bp}=\eta_{12}^{xy}\eta_{\bp1}^{x+y}[x,y],\\
%& g_P[x,x]g^\dagger_P=\eta_{12}^x[x,x],\\
%& g_{\bp}[x,-x]g^\dagger_{\bp}=\eta_{12}^x[x,-x].
%\eea

Once the symmetry transformations of fermionic spinons are determined, the associated mean-field amplitudes $[x,y]\equiv\langle0,0|x,y\rangle$ must obey certain symmetry constraints. For example, on-site chemical potential $\hat{\mu}\equiv[0,0]$ must satisfy
\bea\label{sym cond:onsite}
&g_\bst\hat{\mu}g_\bst^\dagger=-\hat\mu,\\
&\notag \hat\mu=g_\bss\hat\mu g_\bss^\dagger=g_\cs\hat\mu g_\cs^\dagger.
\eea

Mean-field amplitude $\hat\alpha\equiv[1,1]$ between nearest neighbors (NNs) must have
\bea\label{sym cond:1st NN}
&g_\bst\hat{\alpha}g_\bst^\dagger=-\hat\alpha,\\
&\notag \hat\alpha=\eta_{12}g_\bss\hat\alpha g_\bss^\dagger,\\
&\notag\hat\alpha^\dagger=\eta_{12}g_\cs^3\hat\alpha g_\cs^{-3}.
\eea

For next nearest neighbor (NNN) mean-field amplitude $\hat\beta\equiv[2,1]$ we have
\bea\label{sym cond:2nd NN}
&g_\bst\hat{\beta}g_\bst^\dagger=-\hat\beta,\\
&\notag \eta_{12}g_\cs\hat\beta g_\cs^\dagger=g_\bss\hat\beta g_\bss^\dagger,\\
&\notag\hat\beta^\dagger=\eta_{12}g_\cs^3\hat\beta g_\cs^{-3}.
\eea

Among the 20 different Abrikosov fermion states summarized in TABLE \ref{tab:PSG}, only 4 states can realize a symmetric $Z_2$ spin liquid with up to NNN mean-field amplitudes. They are \#1, \#6, \#18 and \#20 as highlighted by red color in TABLE \ref{tab:PSG}. Later we'll show that \#18 state doesn't support a gapped spinon spectrum with up to NNN amplitudes.

\subsection{Duality between Abrikosov-fermion and Schwinger-boson representations}\label{sec:duality}

In Schwinger-boson representation\cite{Auerbach1994B}, a spin-$1/2$ particle on lattice site ${\bf r}$ is decomposed into two species of bosonic spinons $\{b_{{\bf r},\alpha}|\alpha=\uparrow/\downarrow\}$:
\bea
\vec{S}_{\bf r}=\frac12\sum_{\alpha,\beta=\uparrow/\downarrow}b^\dagger_{{\bf r},\alpha}\vec\sigma_{\alpha,\beta} b_{{\bf r},\beta}
\eea
where $\vec\sigma$ are Pauli matrices. Similar to Abrikosov-fermion representation, the spin wavefunction is obtained by single-occupancy Gutzwiller projection on mean-field states of bosonic spinons. Schwinger-boson representation has a $U(1)$ gauge redundancy $b_{{\bf r},\sigma}\longrightarrow e^{\imth\phi_{\bf r}}b_{{\bf r},\sigma}$ and hence any symmetry operation $\hat u$ is followed by a $U(1)$ phase rotation of Schwinger bosons by angle $\phi_{\hat u}(x,y)$. \Ref{Wang2006} classified $Z_2$ spin liquids in Schwinger-boson representation and obtain 8 different states on triangular lattice, with the following gauge-inequivalent PSGs ($p_i=0,1$):
\bea
&\phi_{T_1}(x,y)=0,~~~\phi_{T_2}(x,y)=xp_1\pi,\\
&\notag\phi_\bss(x,y)=(xyp_1+{p_2}/2)\pi,\\
&\notag\phi_\cs(x,y)=\big[p_1y(x+\frac{y-1}2)+{p_3}/2\big]\pi.
\eea
and one can always choose a gauge so that $\phi_\bst(x,y)=0$ for time reversal $\bst$.

\begin{table}[tb!]
\begin{tabular} {|c|c|c|c|c}
\hline
Algebraic Identities&  bosonic $b_{\sigma}$ & fermionic $f_{\sigma}$ & vison $v=b\times f$\\ \hline
$T^{-1}_{2}T^{-1}_{1}T_{2}T_{1}=\bse$&(-1)$^{p_1}$&$\eta_{12}$&-1\\ \hline
$\bss^{-1}T_1\bss T_2^{-1}=\bse$&1& 1&1\\ \hline
$\bss^{-1}T_2\bss T_1^{-1}=\bse$&1&1&1\\ \hline
$\cs^{-1}T_1\cs T_2=\bse$&1&1&1\\ \hline
$\cs^{-1}T_2\cs T_2^{-1}T_1^{-1}=\bse$&1&1&1\\ \hline
$\bss^2=\bse$&(-1)$^{p_2}$&$\eta_\bss$&1\\ \hline
$R^{2}=(\cs\bss)^2=\bse$&(-1)$^{p_2+p_3}$&$\eta_{\bss\cs}$&1\\ \hline
$(\cs)^{6}=\bse$&(-1)$^{p_3}$&$\eta_\cs$&-1\\ \hline
$T_1^{-1}\bst^{-1}T_1\bst=\bse$&1&1&1\\ \hline
$T_2^{-1}\bst^{-1}T_2\bst=\bse$&1&1&1\\ \hline
$\bss^{-1}\bst^{-1}\bss\bst=\bse$&(-1)$^{p_2}$&$\eta_{\bss\bst}$&1\\ \hline
$R^{-1}\bst^{-1}R \bst=\bse$&(-1)$^{p_2+p_3}$&$\eta_{\cs\bst}\eta_{\bss\bst}$&1\\ \hline
$\bst^2=\bse$&-1&-1&1\\ \hline
\end{tabular}
\caption{The correspondence between bosonic spinon\cite{Wang2006}, fermionic spinon and vison PSGs on the triangular lattice. %Given the signs of the algebras of the symmetry operations on spinons $v_{\sigma}$ and $\psi_{\sigma}$, one can deduce\cite{Essin2013,Mesaros2013} those of the algebras of the symmetry operations on the spin-singlet vison operator $w \sim v\times \psi$.
Here bosonic spinon ($b_\sigma$) PSGs are labeled by three integers\cite{Wang2006} $p_i=0,1~(i=1,2,3)$, while fermionic spinon ($f_\sigma$) PSGs are labeled by six integers $(\eta_{12},\eta_\bss,\eta_{\bss\cs},\eta_{\cs},\eta_{\bss\bst},\eta_{\cs\bst})$ where $\eta=\pm1$.
Choosing a proper gauge we can always fix $\cs^{-1}T_2\cs T_2^{-1}T_1^{-1}=\cs^{-1}T_1\cs T_2=1$ for both spinons and visons.}
\label{tab:unification}
\end{table}

What is the relation between these 8 Schwinger-boson states and the 20 Abrikosov-fermion states obtained in this work? There are two ways to achieve this connection. The first approach is to directly compare the crystal symmetry (mirror reflection $\bss,R$ and inversion $(C_6)^3$) quantum numbers of their projected spin wavefunction on different finite lattices\cite{Zaletel2015} and on infinite cylinders\cite{Zaletel2015,Qi2015}. For symmetry quantum numbers associated with time reversal symmetry (3 row at the bottom of TABLE \ref{tab:unification}), one can compare the degeneracy in the entanglement spectra of their projected wavefunctions on different infinite cylinders with $\bss$ and $R$ symmetries\cite{Zaletel2015}. Repeating the same analysis as in \Ref{Zaletel2015} we obtain the unification of Schwinger-boson PSGs and Abrikosov-fermion PSGs as shown in TABLE \ref{tab:PSG}.

The 2nd approach is to understand the vison PSGs in the Schwinger-boson state. As argued in \Ref{Lu2014} any $Z_2$ spin liquids constructed by projecting a mean-field Schwinger-boson state doesn't support gapless edge states\cite{Reuther2014}. On the other hand, nontrivial vison PSGs, \eg $\bss^2=-1$ or $\bss\bst\bss^{-1}\bst^{-1}=-1$ acting on visons $v$, will lead to symmetry protected gapless edge states on an open edge preserving reflection $\bss$. These strong constraints can determine the vison PSGs in any Schwinger-boson state, as shown in the right column of TABLE \ref{tab:unification}. Meanwhile, in a $Z_2$ spin liquid the fermionic spinon $f_\sigma$ can be viewed as a bound state\cite{Kitaev2003} of a bosonic spinon $b_\sigma$ and vison $v$ due to fusion rule $f_\sigma=b_\sigma\times v$. Consequently the fermionic spinon ($f$) PSGs is a product of bosonic spinon ($b$) PSGs and vison ($v$) PSGs, multiplied by a twist factor\cite{Essin2013,Lu2014,Zaletel2015,Qi2015}. This twist factor is -1 in the case of $(\cs)^6=\bse$\cite{Essin2013}, $\bss^2=R^2=\bse$\cite{Lu2014,Zaletel2015,Qi2015} and $\bss\bst\bss^{-1}\bst^{-1}=R\bst R^{-1}\bst^{-1}=\bse$\cite{Zaletel2015,Lu2014}, and +1 in all other cases.

As a result in TABLE \ref{tab:unification}, a Schwinger-boson state $(p_1,p_2,p_3)$ belong to the same symmetric $Z_2$ spin liquid phase as an Abrikosov-fermion state if their PSGs have the following correspondence:
\bea\label{duality:algebra}
&\eta_{12}=(-1)^{p_1},\\
&\notag \eta_{\bss}=\eta_{\bss\bst}=\eta_\cs\eta_{\bss\cs}=(-1)^{p_2+1},\\
&\notag\eta_{\cs}=\eta_{\cs\bst}=(-1)^{p_3}.
\eea
All 8 Schwinger-boson states have their counterparts in Abrikosov-fermion representation, as summarized in TABLE \ref{tab:PSG}. Since the continuous phase transitions between $Z_2$ spin liquids and noncollinear magnetic orders can be simply achieved by condensing bosonic spinons\cite{Sachdev1992,Chubukov1994a} in the Schwinger-boson representation, this unification allows us to understand the neighboring magnetic orders of some Abrikosov-fermion states.

\subsection{Nematic descendants}\label{sec:nematic classification}

DMRG studies\cite{Zhu2015,Hu2015} suggest strong evidence for nematic order in the gapped spin liquid phases, \ie spontaneously broken $C_6$ rotational symmetry coexisting with $Z_2$ topological order. This motivates us to study nematic (or anisotropic) $Z_2$ spin liquids on a triangular lattice. In spite of broken $C_6$ symmetry, these nematic $Z_2$ spin liquids still preserve two mirror reflections\cite{Zhu2015,Hu2015}, $\bss$ and $R\equiv\bss(\cs)^3$, as well as their combination: site-centered inversion symmetry $I=(\cs)^3$. \Ref{Zhou2002} classified nematic $Z_2$ spin liquids on an anisotropic triangular lattice (with $\bss$ and $R$ symmetry), leading to at least 63 distinct Abrikosov-fermion states. Instead of examining all these states, here we focus on those nematic states in proximity to a $C_6$-symmetric $Z_2$ spin liquids. In other words they can be achieved by adding anisotropic perturbations to a $C_6$-symmetric $Z_2$ spin liquid, \ie one of the 20 states in TABLE \ref{tab:PSG}.

It turns out each of the 20 Abrikosov-fermion states summarized in TABLE \ref{tab:PSG} has a unique nematic descendant with the following symmetry transformations
\bea
&\notag G_{T_1}(x,y)=1,~~~G_{T_2}(x,y)=\eta_{12}^x,\\
&\notag G_\bss(x,y)=\eta_{12}^{xy}g_\bss,~~~G_I(x,y)=\eta_{12}^{x+y}(g_\cs)^3,\\
&\notag G_R(x,y)=\eta_{12}^{xy+x+y}g_R,~~~G_\bst(x,y)=g_\bst=\imth\tau^2.
\eea
where $g_R\equiv g_\bss(g_\cs)^3$. These symmetry properties impose the following constraints on a generic mean-field amplitude $[x,y]\equiv\langle0,0|x,y\rangle$:
\bea
&\notag g_\bst[x,y]g^\dagger_\bst=-[x,y],\\
&\notag g_\cs^3[x,y]g_\cs^{-3}=\eta_{12}^{xy+x+y}[x,y]^\dagger,\\
&\notag g_\bss[x,x]g^\dagger_\bss=\eta_{12}^x[x,x],\\
& g_R[x,-x]g^\dagger_R=\eta_{12}^x[x,-x].
\eea
To be specific, there are two independent amplitudes between 1st nearest neighbors (NNs) \ie $[1,0]$ and $[1,1]$ in nematic states. They satisfy
\bea
&\notag g_\bst[1,0]g^\dagger_\bst=-[1,0],\\
&g_\cs^3[1,0]g_\cs^{-3}=\eta_{12}[1,0]^\dagger.
\eea
and
\bea
&\notag g_\bst[1,1]g^\dagger_\bst=-[1,1],\\
&\notag g_\cs^3[1,1]g_\cs^{-3}=\eta_{12}[1,1]^\dagger,\\
& g_\bss[1,1]g^\dagger_\bss=\eta_{12}[1,1].
\eea
Similarly, two independent 2nd NN amplitudes $[2,1]$ and $[1,-1]$ satisfy
\bea
&\notag g_\bst[2,1]g^\dagger_\bst=-[2,1],\\
&g_\cs^3[2,1]g_\cs^{-3}=\eta_{12}[2,1]^\dagger.
\eea
and
\bea
&\notag g_\bst[1,-1]g^\dagger_\bst=-[1,-1],\\
\notag&g_\cs^3[1,-1]g_\cs^{-3}=\eta_{12}[1,-1]^\dagger,\\
& g_\bss[1,-1]g^\dagger_\bss=[1,-1]^\dagger.
\eea

These conditions give rise to symmetry-allowed mean-field amplitudes for the 20 nematic $Z_2$ spin liquids, as summarized in TABLE \ref{tab:PSG}. Due to spontaneous breaking of $C_6$ rotational symmetry, the symmetry conditions for nematic $Z_2$ spin liquids are less stringent than for the isotropic ($C_6$ symmetric) ones, leading to nonzero NN and NNN amplitudes that are not allowed in isotropic states. As a consequence, 4 more states support a nematic $Z_2$ spin liquid with up to NNN mean-field amplitudes, in addition to the 4 isotropic states (red-colored in TABLE \ref{tab:PSG}) in isotropic case. They are \#4, \#7, \#10 and \#13, as highlighted by blue color in TABLE \ref{tab:PSG}.

Previous variational Monte Carlo study\cite{Mishmash2013} suggested a gapless $d$-wave paired state of Abrikosov fermions, which is energetically competitive for a finite range of $J_2/J_1$. In this state the fermionic spinons share the same PSG with state \#13 in TABLE \ref{tab:PSG}.

\section{Gapped symmetric $Z_2$ spin liquids and their neighboring phases}\label{SYMMETRIC Z2SL}

In DMRG studies of spin-$1/2$ $J_1$-$J_2$ Heisenberg model, the gapped spin liquid phase appear only in the parameter range\cite{Zhu2015,Hu2015} of $J_2/J_1\leq0.17$. Therefore for candidate $Z_2$ spin liquids in Abrikosov-fermion representation, we focus on $Z_2$ spin liquid states that can be realized by up to next nearest neighbor (NNN) mean-field amplitudes. In particular we'll start from mean-field ansatz with nonzero nearest neighbor (NN) amplitudes. We can always choose a proper gauge such that NN mean-field amplitudes are all real hoppings ($\tau^3$).

There are only two possible NN hopping ansatz that preserve all symmetries (time reversal, spin rotation and crystal symmetries):

(I) uniform RVB state: all NN real hoppings share the same amplitude $\alpha$.

(II) $\pi$-flux (algebraic spin liquid) state: $\pi$ flux is inserted in half of the triangles (see FIG. \ref{fig:triangle pi-flux}).

They correspond to two distinct symmetric $U(1)$ spin liquids on triangular lattice, which host gapless fermionic spinons coupled to $U(1)$ gauge field. In the following we'll study candidate $Z_2$ spin liquids by adding perturbations to these two parent $U(1)$ spin liquids.

\subsection{Gapped $Z_2$ spin liquids near uniform RVB state}

Among the 20 states in TABLE \ref{tab:PSG}, 4 of them support uniform NN real hoppings, corresponding to 4 symmetric $Z_2$ spin liquids in the neighborhood of uniform RVB state. They are \#1, \#5, \#13 and \#19 states. Three of these 4 states (\#5, \#13 and \#19) only allow real hoppings up to NNN amplitudes, therefore remaining to be gapless $U(1)$ spin liquids with spinon fermi surfaces.

On the other hand, \#1 state allow uniform hoppings and pairings in NN and NNN amplitudes. It's straightforward to verify that uniform pairing terms can open up a gap on the spinon fermi surface of NN hoppings, giving rise to a gapped $Z_2$ spin liquid which preserves all symmetries. Therefore \#1 state, which is dual to Schwinger-boson (1,1,0) state, is the only gapped $Z_2$ spin liquid phase in the neighborhood of uniform RVB state.

\subsection{Gapped $Z_2$ spin liquids near $\pi$-flux algebraic spin liquid}

Unlike the uniform RVB state with a spinon fermi surface, the $\pi$-flux state\cite{Rachel2015} is an algebraic spin liquid\cite{Wen2002,Hermele2005}, described by gapless Dirac fermions coupled to emergent $U(1)$ gauge field. Similar to the case of uniform RVB state, there are 4 distinct $Z_2$ spin liquids in the neighborhood of $\pi$-flux state: \ie \#10, \#12, \#18 and \#20 states in TABLE \ref{tab:PSG}.

Choosing the 4-site magnetic unit cell as illustrated in FIG. \ref{fig:triangle pi-flux}, the Bloch Hamiltonian in the basis of $\psi_{(r_1,r_2),\sigma}\equiv(f_{(2x,2y+1),\sigma},f_{(2x+1,2y+1),\sigma},f_{(2x,2y),\sigma},f_{(2x+1,2y),\sigma})^T$ writes
\bea
&\notag\hat H^{\pi}_{MF}=\alpha\sum_{{\bf k},\sigma}\psi_{{\bf k},\sigma}^\dagger \hat h_{\bf k}\psi_{{\bf k},\sigma},~~~\hat h_{\bf k}=\\
&\notag \bpm
0&1+e^{-\imth k_1}&1+e^{\imth k_2}&e^{\imth k_2}-e^{-\imth k_1}\\
1+e^{\imth k_1}&0&e^{\imth(k_1+k_2)}-1&1+e^{\imth k_2}\\
1+e^{-\imth k_2}&e^{-\imth(k_1+k_2)}-1&0&-1-e^{-\imth k_1}\\
e^{-\imth k_2}-e^{\imth k_1}&1+e^{-\imth k_2}&-1-e^{\imth k_1}&0\epm
\eea
The dispersion crosses zero energy (fermi level) at momentum $k_1=k_2=\pi$. Expanding the above Bloch Hamiltonian around Dirac point $(\pi,\pi)$ leads to the following linearized Dirac Hamiltonian in the basis of $\psi_{{\bf q}+(\pi,\pi),\sigma}$
\bea
&\notag\hat h_{\bf k}=\big[-q_1\mu_z\nu_y+q_2\mu_y\nu_0+(q_1+q_2)\mu_x\nu_y\big]\sigma_0+O(|{\bf q}|^2)\\
&\notag=\sqrt6(q_x\gamma_x+q_y\gamma_y)+O(|{\bf q}|^2),\\
&\notag\gamma_x=\frac{(\mu_x-2\mu_z)\nu_y-\mu_y\nu_0}{\sqrt6},~~~\gamma_y=\frac{\mu_x\nu_y+\mu_y\nu_0}{\sqrt2},\\
&\notag{\bf q}={\bf k}-(\pi,\pi)\equiv(q_1,q_2)=(2q_x,\sqrt3q_y-q_x).
\eea
where $\vec\nu$ and $\vec\mu$ are Pauli matrices for the (diamond,dot) and (red,blue) sublattice indices respectively (see FIG. \ref{fig:triangle pi-flux}), while $\vec\sigma$ are Pauli matrices associated with spin index. We've set lattice constant as unity.

In the Nambu basis of $\Psi_{\bf q}\equiv(\psi^T_{{\bf q},\sigma},-\imth\psi^\dagger_{{-\bf q},\sigma}\sigma_y)^T$, the low-energy Dirac Hamiltonian writes
\bea\label{dirac ham:triangular}
&\hat{H}_{Dirac}^\pi=\frac\alpha2\sum_{\bf q}\Psi^\dagger_{\bf q}\hat d_{\bf q}\Psi_{\bf q},\\
&\notag \hat{d}_{\bf q}=\big[-q_1\mu_z\nu_y+q_2\mu_y\nu_0+(q_1+q_2)\mu_x\nu_y\big]\sigma_0\tau_z\\
&\notag=\sqrt6[q_x\gamma_x+q_y\gamma_y]\sigma_0\tau_z.
\eea
where $\vec\tau$ are Pauli matrices for Nambu index.

A gapped $Z_2$ spin liquid can be obtained by adding pairing terms, which open up a gap in the Dirac spectrum. These pairing ``mass terms'' for Dirac fermions, however, may break certain symmetries. In particular, Dirac fermion $\Psi_{\bf q}=\sigma_y\tau_y\Psi^*_{-\bf q}$ in (\ref{dirac ham:triangular}) transform under symmetries in the following way:
\bea
&\notag\Psi_{\bf q}\overset{\bst}\longrightarrow\imth\sigma_y\Psi_{-\bf q},\\
&\notag\Psi_{\bf q}\overset{SU(2)_{spin}}\longrightarrow\exp(\imth\frac\theta2\hat{n}\cdot\vec\sigma)\Psi_{\bf q},\\
&\notag\Psi_{\bf q}\overset{T_1}\longrightarrow\mu_0\nu_y\sigma_0\tau_z\Psi_{\bf q},\\
&\notag\Psi_{\bf q}\overset{T_2}\longrightarrow\mu_y\nu_z\sigma_0\tau_z\Psi_{\bf q},\\
&\notag\Psi_{(q_1,q_2)}\overset{\bss}\longrightarrow U_\bss\sigma_0(\tilde g_\bss)\Psi_{(q_2,q_1)},\\
&\notag\Psi_{(q_1,q_2)}\overset{R=\bss\cs^3}\longrightarrow U_R\sigma_0(\tilde g_\bss \tilde g_\cs^3)\Psi_{(-q_2,-q_1)},\\
&\notag\Psi_{(q_1,q_2)}\overset{\cs}\longrightarrow U_\cs\sigma_0(\tilde g_\cs)\Psi_{(-q_2,q_1+q_2)}.
\eea
where we define $\tilde{g}_\bss=e^{\imth\pi\tau^2/4}g_\bss e^{-\imth\pi\tau^2/4}$ and $\tilde{g}_\cs=e^{\imth\pi\tau^2/4}g_\cs e^{-\imth\pi\tau^2/4}$ in a different gauge. We have $\{\tilde g_\cs,\tau_z\}=\{\tilde g_\bss,\tau_z\}=0$ and
\bea
&\notag U_\bss=U_R=\bpm0&0&0&1\\0&-1&0&0\\0&0&1&0\\1&0&0&0\epm,~U_\cs=\bpm0&0&0&-1\\1&0&0&0\\0&0&1&0\\0&-1&0&0\epm.
\eea
Notice that there is a $U(1)$ gauge redundancy for algebraic spin liquid (\ref{dirac ham:triangular}), \ie the Dirac Hamiltonian is invariant under any $\tau_z$ rotations.

Clearly the symmetric spin-singlet pairing mass terms which anticommute with Dirac matrices $\gamma_{x,y}\sigma_0\tau_z$ are simply
\bea\label{mass:singlet sc}
{\bf M}_{sSC}=\mu_0\nu_0\sigma_0(\Re\Delta_{ssc}\cdot\tau_x+\Im\Delta_{ssc}\cdot\tau_y).
\eea
Time-reversal-invariant real pairing mass $\mu_0\nu_0\sigma_0\tau_x$ preserves $\bss$ and $\cs$ symmetries, if and only if $[\tilde g_\cs,\tau_x]=[\tilde g_\bss,\tau_x]=0$. Therefore among the 4 neighboring $Z_2$ spin liquids near $\pi$-flux state, only \#20 state with $g_\bss=g_\cs=\imth\tau_z$ allows a symmetric pairing mass for the Dirac fermion.

As a result, \#20 state which is dual to Schwinger-boson $(0,0,1)$ state is the only gapped $Z_2$ spin liquid in the neighborhood of the $\pi$-flux algebraic spin liquid.

\subsection{Confined symmetry-breaking phases in proximity to gapped $Z_2$ spin liquids}

A gapped $Z_2$ spin liquid can be driven into ``confined'' symmetry-breaking phases with no fractional excitations, by condensing their fractional excitations. For example, condensing spin-singlet vison excitations typically gives rise to a valence bond solid (VBS) phase, which breaks crystal symmetries but preserves global spin rotation and time reversal symmetries. On the other hand, condensing bosonic spinons (carrying spin-$1/2$ each) typically leads to a noncolinear magnetic order which breaks time reversal as well as spin rotational symmetries. Therefore confined symmetry-breaking states in proximity to a $Z_2$ spin liquid reflect the symmetry quantum numbers of spinons and visons therein, and is an important characterization of the symmetry fractionalization in the $Z_2$ spin liquid. In the following we analyze the proximate ordered phases of the two promising $Z_2$ spin liquids, \#1 and \#20 as discussed above.

First we discuss \#1 state in the neighborhood of uniform RVB state. As shown in section \ref{sec:duality}, it describes the same $Z_2$ spin liquid phase as the (1,1,0) state in Schwinger-boson representation\cite{Wang2006}. In the Schwinger-boson representation, the transition from $Z_2$ spin liquid to a magnetic ordered phase is simply described by condensation of Schwinger-bosons. As shown in \Ref{Wang2006}, the consequent ordered phase in proximity to (1,1,0) state is a 4-sublattice noncolinear magnetic order:
\bea
\notag S({\bf r})={\bf n}_1(-1)^x+{\bf n}_2(-1)^y+{\bf n}_3(-1)^{x+y}.
\eea
It features a $2\times2$ magnetic unit cell and is the tetrahedral state classified in \Ref{Messio2011}.

On the other hand, the low-energy dynamics of spinless visons is described by a fully frustrated quantum Ising model on the dual honeycomb lattice\cite{Moessner2000,Moessner2001b}, which is constrained by the vison PSGs in TABLE \ref{tab:unification}. As shown in \Ref{Moessner2000,Moessner2001b,Misguich2008,Slagle2014}, the VBS phase achieved by condensing visons in the $Z_2$ spin liquid is featured by an enlarged $\sqrt{12}\times\sqrt{12}$ unit cell.

Now let's turn to \#20 state in the neighborhood of $\pi$-flux algebraic spin liquid. It belongs to the same $Z_2$ spin liquid phase as (001) state in Schwinger-boson representation\cite{Sachdev1992,Wang2006}. Since all $Z_2$ spin liquids constructed from Schwinger-boson mean-field ansatz share the same vison PSGs\cite{Lu2014}, the VBS phase in proximity to \#20 state generically has the same $\sqrt{12}\times\sqrt{12}$ symmetry-breaking pattern as the \#1 state. Meanwhile upon condensation of bosonic spinons, \Ref{Sachdev1992,Wang2006} showed that the so-called 120-degree Neel ordered phase with a $\sqrt3\times\sqrt3$ magnetic unit cell will be developed from the (001) state.

Since \#20 state can be obtained by adding a singlet pairing mass term to Dirac spinons in $\pi$-flux state, the Abrikosov-fermion representation provides an alternative way to understand the confined symmetry-breaking phases in its neighborhood. To be specific, the VBS phase corresponds to the following mass term added to Dirac spinons in (\ref{dirac ham:triangular}):
\bea\label{mass:vbs}
&{\bf M}_{VBS}=\vec B\cdot{\bf V}=\frac1{\sqrt3}\Big[B_1(\mu_x+\mu_z-\mu_y\nu_y)+\\
&\notag B_2(\nu_x+\mu_x\nu_z-\mu_z\nu_z)+B_3(\nu_z-\mu_x\nu_x+\mu_z\nu_x)\Big]\tau_z
\eea
where we've only written down the Pauli matrices with a nonzero subscript. The VBS mass matrices ${\bf V}$ are invariant under time reversal and spin rotations, but transform nontrivially under space group symmetries:
\bea
&\notag T_1(V_1,V_2,V_3)T_1^{-1}=(V_1,-V_2,-V_2),\\
&\notag T_2(V_1,V_2,V_3)T_2^{-1}=(-V_1,-V_2,V_2),\\
&\notag \bss(V_1,V_2,V_3)\bss^{-1}=(-V_3,V_2,-V_1),\\
&\notag \cs(V_1,V_2,V_3)\cs^{-1}=(-V_2,V_3,-V_1).
\eea
corresponding to crystal-symmetry-breaking VBS orders. It's straightforward to verify that 3 VBS masses $V_{1,2,3}$ anticomute with each other, and they all anticommute with the singlet pairing mass $\tau_{x,y}$ corresponding to $Z_2$ spin liquid phase. As a result the low-energy dynamics of 5-component mass-term vector
\bea
&\notag\vec v=(\Re\Delta_{ssc},\Im\Delta_{ssc},VBS_1,VBS_2,VBS_3)
\eea
is described by $O(5)$ non-linear sigma model (NLsM) with a Wess-Zumino-Witten (WZW) term\cite{Abanov2000,Tanaka2005,Senthil2006}. Such a WZW term can be derived by integrating out fermionic spinons in the Dirac Hamiltonian (\ref{dirac ham:triangular}). It implies a continuous quantum phase transition\cite{Lu2011a,Lu2014} between $Z_2$ spin liquid \#20 ($\Delta_{ssc}\neq0,~VBS=0$) and the VBS phase ($\Delta_{ssc}=0,~VBS\neq0$).

Similarly the quantum spin Hall (QSH) mass terms to Dirac spinons
\bea\label{mass:qsh}
{\bf M}_{QSH}=\vec n\cdot{\bf m}=\frac{(\mu_x\nu_y+\mu_z\nu_y-\mu_y)(\vec n\cdot\vec\sigma)\tau_z}{\sqrt3}
\eea
also anticommute with the singlet pairing masses. This indicates the dynamics of the following 5-component mass vector
\bea
&\notag\vec v^\prime=(\Re\Delta_{ssc},\Im\Delta_{ssc},QSH_x,QSH_y,QSH_z)
\eea
is also captured by $O(5)$ NLsM with a WZW term. Once the Dirac spinons acquires a finite gap from the QSH mass term, the monopole of $U(1)$ gauge fields coupled to fermionic spinons will carry a spin quantum number due to QSH effect\cite{Ran2008,Lu2011a,Lu2014}. Although QSH mass ${\bf M}_{QSH}$ in (\ref{mass:qsh}) appear to preserve $U(1)$ spin rotation along $\vec n$-axis, the proliferation of monopole events in $U(1)$ gauge fluctuations will destroy the spin conservation along $\hat n$-axis, leading to a noncolinear magnetic order with 3 Goldstone modes\cite{Lu2011a,Lu2014}. Therefore the WZW term for the quituplet of singlet pairing and QSH masses implies a continuous quantum phase transition between 120-degree Neel order and $Z_2$ spin liquid \#20, in accordance with Schwinger-boson (001) state\cite{Sachdev1992,Wang2006}.

In the phase diagram of triangular $J_1$-$J_2$ Heisenberg model obtained in DMRG studies, the gapped spin liquid phase appear in proximity to 120-degree Neel order and another colinear ``stripy'' Neel order\cite{Zhu2015,Hu2015}. This is another evidence in favor of \#20 state (\ie Schwinger-boson (0,0,1) state) as a promising candidate for the spin liquid found in $J_1$-$J_2$ model.

\begin{figure}
\includegraphics[width=0.8\columnwidth]{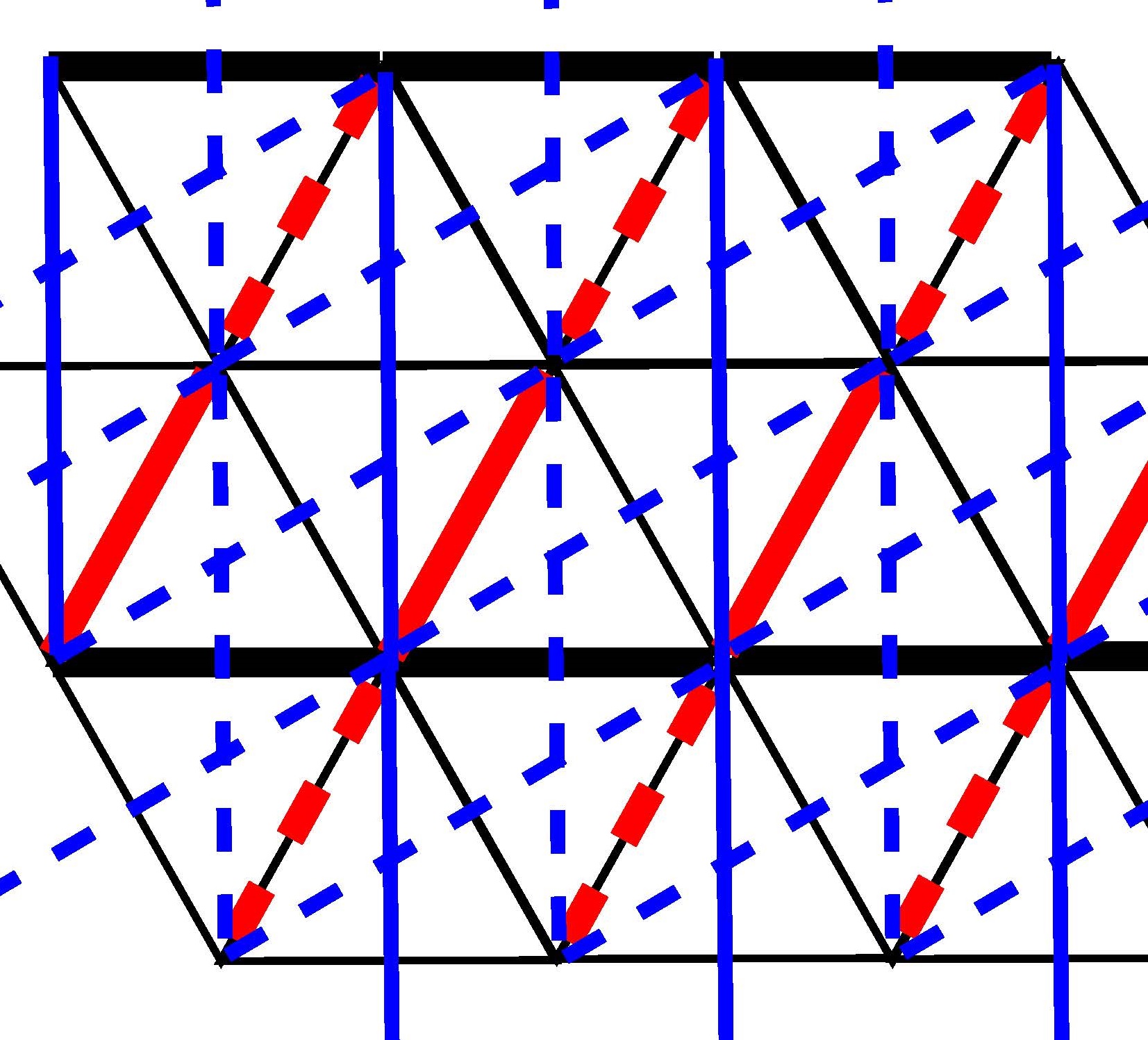}
\caption{(color online) Mean-field hopping amplitudes (up to NNN) of nematic $Z_2$ spin liquid \#20 in TABLE \ref{tab:PSG}. There are two types of NN hoppings denoted by black and red colors: for black ones, thick and thin lines have the same amplitude $|\alpha_1|$ but opposite signs; for red ones, solid and dashed lines have the same amplitude $|\alpha_2|$ but opposite signs. NNN amplitudes are labeled by blue color: solid and dashed lines have the same amplitude $|\beta_1|$ but opposite sings. Onsite real paring is also allowed by symmetry.}
\label{fig:20 nematic}
\end{figure}

\begin{figure}
\includegraphics[width=0.8\columnwidth]{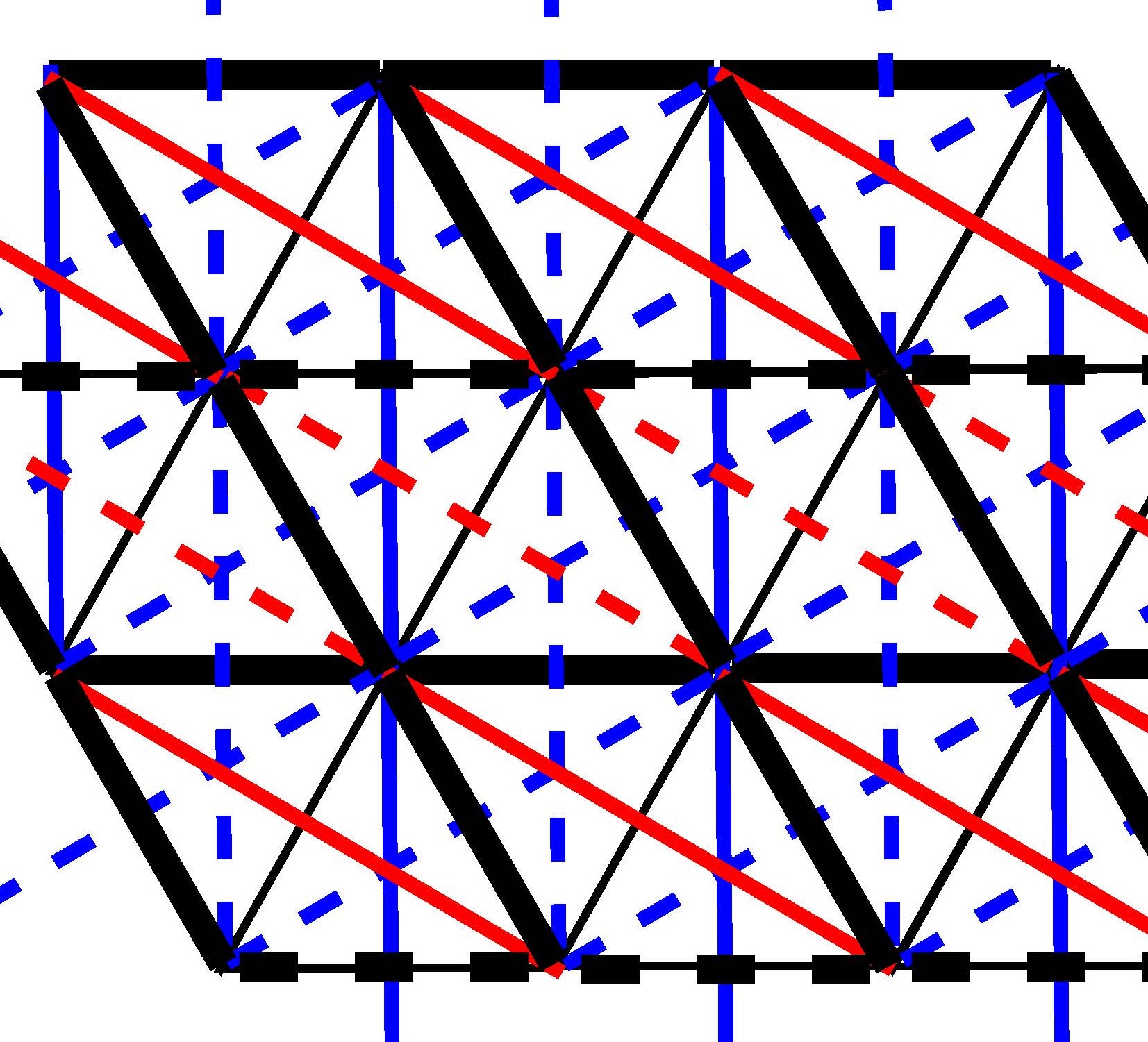}
\caption{(color online) Mean-field hopping amplitudes (up to NNN) of nematic $Z_2$ spin liquid \#6 in TABLE \ref{tab:PSG}. Nonzero NN hoppings are denoted by black thick lines: for black ones, solid and dashed lines have the same amplitude $|\alpha_1|$ but opposite signs. The thin lines connecting NNs have zero hopping amplitude. There are two types of NNN amplitudes: for blue (or red) colors, solid and dashed lines have the same amplitude $|\beta_1|$ (or $|\beta_2|$) but opposite sings. Onsite real paring is also allowed by symmetry.}
\label{fig:6 nematic}
\end{figure}

\section{Nematic $Z_2$ spin liquids on triangular lattice}\label{NEMATIC Z2SL}

In DMRG studies of spin-$1/2$ $J_1$-$J_2$ Heisenberg model on triangular lattice, both the even and odd sectors of the gapped spin liquid phase show spatially anisotropic spin-spin correlations\cite{Zhu2015,Hu2015}, different from the $Z_2$ spin liquid in kagome Heisenberg model\cite{Zhu2015}. Meanwhile the spin liquid shows strong response to a $C_6$-breaking perturbation to Heisenberg couplings\cite{Zhu2015}. This motivates us to seek for nematic $Z_2$ spin liquid candidates for $J_1$-$J_2$ Heisenberg model. Again we focus on those nematic $Z_2$ spin liquids that can be realized by up to NNN mean-field amplitudes.

Nematic $Z_2$ spin liquids on triangular lattice has been classified in \Ref{Zhou2002}. Here we restrict ourselves to those nematic (anisotropic) $Z_2$ spin liquids that can be obtained by perturbing fully-symmetric (isotropic) $Z_2$ spin liquids. As discussed in section \ref{sec:nematic classification}, for each symmetric $Z_2$ spin liquids on triangular lattice, there is one and only one nematic ``descendant'' which preserves two mirror reflections $\bss$ and $R$ but breaks $C_6$ rotation (see FIG. \ref{fig:triangle pi-flux}). Their symmetry-allowed mean-field amplitudes are summarized in TABLE \ref{tab:PSG}.

First we examine those nematic $Z_2$ spin liquids in the neighborhood of uniform RVB state, \#1, \#5, \#13 and \#19. In addition to \#1 state which already realizes a gapped $Z_2$ spin liquid without breaking $C_6$ symmetry, in the presence of nematic order \#13 state can also realizes a $Z_2$ spin liquid with up to NNN amplitudes. To be precise, $C_6$ symmetry breaking gives rise to nonzero pairing amplitudes ($\tau^1$) in \#13 state, as shown in TABLE \ref{tab:PSG}. However, these nematic pairing terms generally cannot open a full gap on the spinon fermi surface: typically they leave 4 point nodes on the fermi surface similar to the d-wave superconductors. Therefore even in the presence of nematic order, \#1 state is still the only gapped $Z_2$ spin liquid near uniform RVB state.

Next we look into nematic $Z_2$ spin liquids in the neighborhood of $\pi$-flux state, \#10, \#12, \#18 and \#20. Among them, only ansatz \#20 supports a fully-symmetric gapped $Z_2$ spin liquid. After breaking $C_6$ symmetry with nematic order, it is possible to realize a $Z_2$ spin liquid with up to NNN amplitudes in both \#10 and \#18 (but not \#12) state. However, the pairing terms cannot open up a gap in the spinon Bogoliubov spectrum: point nodes still exist in \#10 and \#18 state. As a result, \#20 is again the only gapped nematic $Z_2$ spin liquid near $\pi$-flux algebraic spin liquid. In addition to onsite real pairing, it also allows NN/NNN hoppings as demonstrated in FIG. \ref{fig:20 nematic}.

Meanwhile, there are 3 other nematic $Z_2$ spin liquids that can be realized with up to NNN amplitudes: \#4, \#6 and \#7. Among them, \#4 and \#7 have gapless excitations in their spinon Bogoliubov spectra, while \#6 supports a gapped nematic $Z_2$ spin liquid. The mean-field ansatz \#6 allows onsite real hoppings, together with NN/NNN real hoppings as demonstrated in FIG. \ref{fig:6 nematic}.\\

To summarize, there are 3 gapped nematic $Z_2$ spin liquids that can be realized by up to NNN mean-field amplitudes among the 20 states in TABLE \ref{tab:PSG}: they are \#1 state dual to Schwinger-boson (110) state, \#6 state, and \#20 state dual to Schwinger-boson (001) state. They are the most promising candidate for possible nematic $Z_2$ spin liquid phase in triangular-lattice $J_1$-$J_2$ Heisenberg model.

\section{Summary}\label{SUMMARY}

To summarize, in this work we studied symmetry $Z_2$ spin liquids on triangular lattice and their nematic descendants, in connection to the numeric discovery of a gapped spin liquid phase in triangular lattice $J_1$-$J_2$ model. We classified symmetric $Z_2$ spin liquids in the Abrikosov-fermion representation, found 20 distinct states and identified 2 states (\#1 and \#20 in TABLE \ref{tab:PSG}) as the promising candidates which support a gapped $Z_2$ spin liquid with up to NNN mean-field amplitudes. By establishing the correspondence between symmetric Abrikosov-fermion and Schwinger-boson states on triangular lattice, we are able to understand the noncollinear magnetic order phases and VBS phases in the neighborhood of these 2 gapped spin liquids.
In particular state \#20 in TABLE \ref{tab:PSG}, dual to the (001) state in Schwinger-boson representation, is continuously connected to the 120-degree Neel order, which is found to lie in proximity to gapped spin liquid in the DMRG phase diagram\cite{Zhu2015,Hu2015}. Motivated by numerical evidence of possible nematic order in this spin liquid phase\cite{Zhu2015,Hu2015}, we also studied nematic $Z_2$ spin liquids by perturbing the symmetric ones. We identified 3 promising nematic states, \#1, \#6 and \#20 in TABLE \ref{tab:PSG} which can realize a gapped $Z_2$ spin liquid with up to NNN amplitudes. It'll be interesting for future variational Monte Carlo studies to see whether these nematic $Z_2$ spin liquids are energetically more favorable than fully-symmetric states or not.\\

{Note added}: Upon completion of this paper, we became aware of an independent work\cite{Zheng2015} which also classified symmetric $Z_2$ spin liquids in the Abrikosov-fermion representation on triangular lattice.

\acknowledgements

I'm grateful to Zhenyue Zhu and Donna D. N. Sheng for helpful discussions, and especially to Biao Huang and Nandini Trivedi for collaborations on related topics. I thank the hospitality of Tao Li at Renmin University and the Aspen Center for Physics, where part of this work was finished. This work is supported by startup fund at Ohio State University and in part by NSF Grant No. PHYS-1066293 (YML).

%
%\bibliographystyle{C://Users//Yuan-Ming//Dropbox//notes//apsrev_nurl}
%\bibliography{C://Users//Yuan-Ming//Dropbox//notes//bibs}
%

\end{document}